\documentstyle[12pt,aasms4]{article}

\received{21 December 1995}
\accepted{1 February 1996}

\begin{document}

\title{Quasar-Cluster Associations and Gravitational Lensing 
	by Large-Scale Matter Clumps}

\author{Xiang-Ping Wu}
\affil{Department of Physics, University of Arizona, Tucson, AZ 85721 and \\
       Beijing Astronomical Observatory, Chinese Academy of Sciences,
       Beijing 100080, China}

\and

\author{Li-Zhi Fang}
\affil{Department of Physics, University of Arizona, Tucson, AZ 85721}

\begin{abstract}
Motivated by the significant overdensity of background bright quasars 
recently detected behind the foreground clusters of galaxies on scale 
of $10$ arcminutes, we have investigated the possibility of attributing 
the quasar-cluster associations to gravitational lensing by large-scale 
matter inhomogeneities. Based on the conventional lensing models, 
we have shown that the reported quasar overdensity is unlikely to be 
generated by cluster matter alone. The situation does not change 
even if all the clusters of galaxies which follow their spatial two-point 
correlation function are taken into account, while matter
clumps on scale of $>20$ Mpc are also found to be unable to provide
the required mass surface density since their density contrast is
strictly limited by  the anisotropy measurements
of the cosmic background radiation. 
Moreover, we have pointed out that the influence of  a nonzero 
cosmological constant on the quasar-cluster associations is very minor. 
We  conclude that either
the observed quasar number counts have been seriously contaminated
by the magnification bias of matter inhomogeneities of the universe or
there should exists some intercluster matter on scale of 
less than $\sim20$ Mpc, e.g. from cluster-galaxy correlation, 
whose mean cosmic density 
is about an order of magnitude higher than that of clusters of galaxies.
\end{abstract}


\section{Introduction}

A significant overdensity of background bright optical/radio quasars
have been recently detected on scale of $10^{\prime}$ around 
foreground Zwicky/Abell clusters (Rodrigues-Williams \& Hogan 1994; 
Wu \& Han 1995; Rodrigues-Williams \& Hawkins 1995; Seitz \& 
Schneider 1995). Although these unusual quasar-cluster associations 
are generally believed to be the result of statistical lensing of
quasars by foreground gravitational potential, cluster matter alone
is far from explaining the observed amplitudes of the 
quasar overdensity behind clusters.  It then seems that the matter
inhomogeneities on even larger scale ($>10$ Mpc) traced by galaxy
clusters should be taken into account in the explanation of
the reported quasar-cluster associations.

While the overdensity of background quasars on the similar scale 
around foreground galaxies was detected a few year ago (Fugmann 1988;1990; 
Bartelmann \& Schneider 1993b;1994),  the large-scale matter clumps 
that galaxies are associated were also advocated in order to 
produce the quasar-galaxy associations.
Using N-body simulations of galaxies formation, 
Bartelmann \& Schneider (1993a) did find a correlation of high
redshift quasars with low redshift galaxies in the scenario of
magnification bias by the matter of galaxies and their surrounding
large scale structures.   Their results indicate that
galaxies, and probably clusters of galaxies, contribute a minor 
effect on the quasar overdensity on scale of arcminutes. 
Interestingly, the recent work by
Wu, Zhu \& Fang (1995) shows that even on small scale of arcseconds 
the quasar-galaxy associations are actually generated mainly by  
the cluster matter rather than the galaxies.

Compared to the quasar-galaxy associations,
the quasar-cluster associations deal with the matter distributions 
on scale of ranging from $\sim1$ to $\sim10$ Mpc, 
on which galaxy clusters are
strongly correlated, revealed by their two-point correlation function
$\xi(r)$. It is timely and necessary to address the following
question: Is galaxy cluster clustering described by $\xi(r)$ able to 
provide enough gravitational matter to act as lens for the
reported quasar-cluster associations ? A definite answer to such
a question today relies on  the numerical study of various 
models of formation of large-scale structure of the universe. 
However, an analytic investigation,
as we will make in this letter, of the quasar-cluster associations 
in the scenario of gravitational lensing by various matter clumps
may supply us with a very useful clue to the matter distribution 
on large-scale of the universe.

\section{Quasar enhancement factor}

The overdensity of background quasars at an angular distant $\theta$ 
around a foreground galaxy cluster is
described by the enhancement factor $q(\theta)$ (Narayan 1989)
\begin{equation}
q(\theta)=\frac{N[<m+2.5\log \mu(\theta)]}{N(<m)}\frac{1}{\mu(\theta)}
         =\frac{N[>S/\mu(\theta)]}{N(>S)}\frac{1}{\mu(\theta)},
\end{equation}
where $N$ are the intrinsic quasar number counts above a 
limiting magnitude $m$ or a  flux threshold $S$ and $\mu(\theta)$
is the lensing magnification introduced by the foreground 
gravitational potential.
This equation accounts for both the magnification effect 
($2.5\log \mu$ or $S/\mu$) and the area distortion ($1/\mu)$ due to
light deflection by foreground matter.
To compare with the measurements of quasar-cluster associations that
search for quasar number excess over a range with radius $\theta$ around the
cluster center, the average enhancement factor $\langle q(\theta)\rangle$
is employed: $\langle q(\theta)\rangle=2\int_0^{\theta}\;
	q(\theta)\theta d\theta/\theta^2$.

If we assume that the observed quasar number counts are not significantly
contaminated by  gravitational lensing due to the matter clumps in
the universe, then the number-magnitude relation $N(<B)$
from Boyle, Shanks \& Peterson (1988)  and the
source counts $N(>S)$ at 5 GHz from Langston et al. (1990) 
can be adopted for the optically-selected and the radio-selected quasars, 
respectively.  However, it should be noted that 
the radio counts $N(>S)$ contain both quasars and galaxies and the fraction
of quasars in $N(>S)$ varies with the flux threshold. Therefore,
the employment of $N(>S)$ in the study of quasar-cluster associations 
can only be regarded as an approximate estimate of $\langle q\rangle$.  
Furthermore, quasars are treated as pointlike sources, which would be
suitable for clusters and large-scale matter clumps as lenses.

The observational data of the four searches for quasar-cluster 
associations provide actually the variations of $\langle q\rangle$ 
with the search distance and/or the limiting magnitude.
We don't intend to fit the curves from the theoretical modeling 
of lensing systems as did by Wu \& Han
(1995) and Rodrigues-Williams \& Hawkins (1995). Instead, 
we adopt the one significant result of $\langle q\rangle$ measured 
at a fixed $\theta$ and a limiting magnitude (or flux).  
Rodrigues-Williams \& Hogan (1994) have explicitly  given 
the enhancement factor within 6 Zwicky radii and 
$B<18.5$. The enhancement of $\langle q\rangle$ versus $\theta$ 
becomes nearly unity at $\theta>5^{\prime}$ in Wu \& Han (1995), and
therefore, $\theta=4^{\prime}$ seems to be a reasonable ``edge''
for their sample.  
Rodrigues-Williams \& Hawkins (1995) show the variations of 
$\langle q\rangle$ against the quasar limiting magnitude for 
$\theta\approx0.12^{o}$. We take the most significant value at
$B=18$. Seitz \& Schneider (1995) choose the quasars from  
the 1 Jy radio source sample at 5 GHz but actually use the optically
identified sources which are mostly quasars or BL Lac objects at
high redshift $z_s>0.5$. Their most significant signal appears at
$B\leq19$ and $z_s\approx1$. We adopt the value of 
$\langle q\rangle\approx1.3$ at 9 Zwicky radii.
In particular, we keep both the limiting magnitude and the radio flux
threshold, which helps to finger out the reliability of using 
radio source counts $N(>S)$ in the evaluation of $\langle q\rangle$. 
Unfortunately, we cannot read out the uncertainties from their 
data but learn that the result has a very high significance of up to
$98\%$. Table 1 summarizes these four measurements,  
in which the errorbars in $\langle q\rangle_{obs}$ 
are the $1\sigma\sqrt{N}$ errors arising from
 the estimates of both quasar density over the association
area and the mean quasar density.

\section{Modeling of quasar overdensity behind clusters}

Now we work with the lensing models of the matter 
inhomogeneities associated with the foreground clusters and
test what would be required to explain the reported quasar-cluster
associations.  We adopt a flat cosmological model 
with a cosmological constant $\lambda_0$, i.e.,
$\Omega_0+\lambda_0=1$,
 and  $H_0=50$ $h_{50}$ km s$^{-1}$ Mpc$^{-1}$.

Clusters of galaxies were naturally 
thought to be the deflectors for the observed quasar-cluster associations.  
We can evaluate the contribution to $\langle q\rangle$ from clusters by
modeling the cluster matter as a singular isothermal sphere,
which is characterized uniquely by its one-dimension 
velocity dispersion $\sigma_v$.
This profile is the simplest but more or less reasonable model for
dark matter distribution in clusters of galaxies. We count both
the primary quasar images and secondary ones, if any, which are 
gravitationally magnified by
a factor of $\mu(\theta)=|1-\theta_E/\theta|^{-1}$, where
the Einstein radius is $\theta_E=4\pi(\sigma_v/c)^2 D_{ds}/D_s$,
and we use 
$D_{d}$, $D_{s}$ and $D_{ds}$ to denote the angular diameter
distances to foreground lenses, to background sources and from
lenses to sources, respectively. In Table 1
we  list  the cluster velocity dispersion which is 
required to produce the observed enhancement for each of the four
measurements. Apparently, the resulting $\sigma_v$ is substantially
larger than any realistic values for clusters of galaxies. Taking
the mean cluster velocity dispersion as 1000 km s$^{-1}$, we can
estimate that the gravitational mass ($M\sim\sigma_v^2$)  
responsible for the 
quasar-cluster associations is an order of magnitude higher than 
the presently known total cluster mass.  Meanwhile, it is seen that
the real matter clumps which generate the quasar-cluster associations
must deviate from the $r^{-2}$ distribution since one cannot use a single
velocity dispersion parameter to reproduce all the observed 
$\langle q\rangle$.

If the large-scale
matter inhomogeneities traced by clusters of galaxies contribute an 
additional mean surface mass density $\Sigma$ to cluster matter,
the Einstein radii $\theta_E$ of the background quasars
will be increased by a factor of $(1-\Sigma/\Sigma_{crit})^{-1}$,
where $\Sigma_{crit}=(c^2/4\pi G)(D_s/D_dD_{ds})$ is the critical
surface mass density (Turner, Ostriker \&
Gott 1984). The image magnification now reads 
$\mu(\theta)=|1-\theta_E/\theta|^{-1}(1-\Sigma/\Sigma_{crit})^{-2}$.
Therefore, if the large-scale matter inhomogeneities have a 
mass density comparable to the critical one, the magnification factor
can be greatly enhanced. Alternatively, $\Sigma_{crit}$ would be
smaller in a cosmological constant dominated universe than in
the matter-dominated universe, i.e., the same uniform matter sheet would 
act as more efficient lens in a $\lambda_0$ dominated universe
(Wu et al. 1995). Table 1 gives the mean surface mass density
$\Sigma$ of the large-scale inhomogeneities that are needed to explain the
quasar overdensity around clusters.  Recall that the surface mass density
at the cluster center with core radius of $r_c$ is 
$0.087(\sigma_v/10^3 \;{\rm km\; s}^{-1})^2$ $(r_c/0.25\; 
{\rm Mpc})^{-1}$ $h_{50}$ g cm$^{-2}$
and the minimum critical density for a source at $z_s=2$ is 
$\Sigma_{crit}=(0.41,0.28)$ $h_{50}$ g cm$^{-2}$ 
for $\Omega_0=(1.0,0.2)$. Thus, 
the large-scale matter clumps should have their surface mass density
comparable to the one at the cluster center in order to
act as the lenses for the observed quasar enhancement around clusters 
It is noticed that $\Sigma$ deduced from the radio
bright quasar associations with clusters is a factor of $\sim2$ 
larger than the one from the optical quasar samples. As we have 
mentioned before, this is due to the contamination of radio galaxies
in the radio source counts $N(>S)$ we adopted. The data of Seitz \&
Schneider (1995) illustrate very well this effect: The optical quasar 
number-magnitude relation $N(<B)$ results in a $\Sigma$ that 
is a factor of about 2 smaller than the value given by 
the radio source  counts $N(>S)$ for the same set of quasar data.

\placetable{table-1}

In summary, the $r^{-2}$ mass distributions
for clusters of galaxies fail in reproduction of the observed
quasar-cluster associations, no matter how massive they would be.
A uniform surface matter sheet is found to be an accepted model 
as long as its surface density reaches a value comparable to
the one at the cluster center. The introduction of the cosmological
constant does not significantly reduce the demand for such a high
density inhomogeneity on large-scale.

\section{Contributions of large-scale matter clumps}

We now discuss the mass contributions from the cluster-cluster correlation.
The matter clustering that clusters of galaxies trace on large scale
can be quantitatively described by the cluster spatial two-point 
correlation  function $\xi(r)=(r/r_{cc})^{-1.8}$, where the 
correlation length is $r_{cc}\approx40$ Mpc $h_{50}^{-1}$ 
(Postman, Huchra \& Geller 1992).  Since $\xi(r)$ diverges at $r=0$, 
we truncate $\xi(r)$ when $r<r_0$. The probability of finding a 
cluster in the surface element $2\pi \zeta d\zeta$  
at distance $\zeta$ from a cluster on the plane perpendicular to
the line of sight is
\begin{equation}
dP(\zeta)=
4\pi n\zeta d\zeta \;
          \int_{\zeta}^{\infty}[1+\xi(r)(1+z_d)^{\epsilon}]
			\frac{rdr}{\sqrt{r^2-\zeta^2}}, 
\end{equation}
in which $n$ is the mean cluster number density and
$\epsilon$ accounts for the evolution of $\xi(r)$.
The expected mass contribution from all the clusters  following 
$\xi(r)$ can be computed by
the integration of $m(\zeta)dP(\zeta)$ over $\zeta$ from 0 to $\infty$.
Here $m(\zeta)$ is the cluster mass within $d\zeta$ of $\zeta$. 

We again adopt a singular isothermal sphere model for individual
cluster with velocity dispersion $\sigma_v$. Moreover, the cluster
matter distribution is truncated at the cluster gravitational
radius $R_c$ so that the cluster mass is $M_c=2\sigma_v^2R_c/G$.
We consider only those excess population relative to the 
``background'' cluster of mean density $n$, i.e., we take out the
factor ``1'' in $[1+\xi(r)]$. 
The expected mean surface mass density over the area $\pi \zeta_0^2$
provided by clusters with mass $M_c$ is 
$[\int m(\zeta)dP(\zeta)]/(\pi \zeta^2)$, which reads
\begin{equation}
\Sigma(\zeta_0)=4nM_cr_{cc}(1+z_d)^{\epsilon}
\;F(\zeta_0,r_{cc},R_c,r_0),
\end{equation}
where
\begin{equation}
F=\left(\frac{r_{cc}}{\zeta_0}\right)^{0.8}\;
  \int_{0}^{Rc+\zeta_0}\left(\frac{m(\zeta)}{M_c}\right)\;k(r_0,\zeta)
                     \left(\frac{\zeta}{\zeta_0}\right)^{0.2}
                     d\left(\frac{\zeta}{\zeta_0}\right),
\end{equation}
and
\begin{equation}
k(r_0,\zeta)=\left\{
\begin{array}{ll}
\int_{r_0/\zeta}^{\infty}\frac{dx}{x^{0.8}\sqrt{x^2-1}}, 
         \;\;\;\;\;& \zeta<r_0;\\
1.84,\;          & \zeta\geq r_0.
\end{array}\right.
\end{equation}
Summing up the contributions from 
all kinds of clusters with different mass and number density
gives rise to the expected mass surface density: 
$\overline{\Sigma}(\zeta_0)=
4\Omega_c\rho_0(1+z_d)^{\epsilon}
r_{cc}\;F$,
in which $\Omega_c$ represents the fraction of the total cluster matter
in the matter (with the critical mass density $\rho_0$) of the universe.
For the typical cluster radii of $R_c=3$ -- $5$ $h_{50}^{-1}$ Mpc and
the smallest cluster separation of $r_0=5$ -- $10$ $h_{50}^{-1}$ Mpc,
numerical computation shows that $F\approx2\sim3$  over the range of 
$\zeta=1$ -- $20$ $h_{50}^{-1}$ Mpc 
which is comparable to the search distances
from cluster centers in the measurements of quasar-cluster associations. 
Therefore,
\begin{equation}
\overline{\Sigma}=0.01\;\Omega_c\;
	     \left(\frac{1+z_d}{1.15}\right)^{0.8}\;
             \left(\frac{F}{3}\right)\;h_{50}\; {\rm g\;cm}^{-2},
\end{equation}
where we have assumed a stable clustering model $\epsilon=-1.2$ and 
converted the comoving surface mass density into a physical one
by multiplying a factor of $(1+z_d)^2$. These two factors
do not significantly alter our following result since clusters involved 
are at relatively low redshift.
It appears that even if we take $\Omega_c=\Omega_0=1$, the cluster matter
provided by the cluster-cluster correlation is
of an order of magnitude lower than the surface mass density required to
produce the observed overdensity of background quasars around foreground
clusters.

The mass surface density from matter inhomogeneities on 
scale of larger than the coherence length ($\sim50$ Mpc)
of the cluster-cluster correlation can be estimated though
\begin{equation}
\Sigma=\int_0^{\infty}[\rho(r)-\rho_0]\;dr\sim \rho_0\delta R
=1.45\times10^{-3}\delta\left(\frac{R}{100 {\rm Mpc}}\right)
\;h_{50}\;{\rm g\;cm^{-2}},
\end{equation}
where $\delta$ is the mean present density contrast over scale of $R$. 
However, the evaluation of $\delta$ on scale of larger than 
$\sim10$ $h_{50}^{-1}$ Mpc is sharply constrained by
the measurements of temperature fluctuation $\Delta T/T$
of the cosmic background radiation on various angular scales. 
Using the simple model for a spherical density perturbation 
(Fang \& Wu 1993), we can set an upper limit on $\Sigma$ in terms of 
the recent results of $\Delta T/T$. 
It turns out that the resulting $\Sigma$ from any mass clumps on scale of
greater than $\sim20$ $h_{50}^{-1}$ Mpc 
is at least an order of magnitude smaller than 
the mass surface density required to explain 
the quasar-cluster associations. Therefore, it is unlikely that
the observed quasar-cluster associations can be attributed to 
the lensing effect by large-scale ($>20$ $h_{50}^{-1}$ Mpc) 
structures of the universe.

\section{Discussion and conclusions}

We have shown that the strong associations of background quasars
with foreground clusters on scale of $\sim10$ arcminutes cannot
be interpreted as the statistical lensing by  
clusters of galaxies. The situation   
does not improve even when all the cluster matter that follow
the two-point cluster-cluster
correlation function is involved, while the matter
contribution from large-scale structures ($>20$ $h_{50}^{-1}$ Mpc)
is strongly constrained by
the measurements of the temperature anisotropies of the cosmic
background radiation. We are limited to very few 
possibilities to solve the puzzle of quasar-cluster associations. 

An intuitive speculation is  that the reported quasar-cluster 
associations are statistical variations arising from the 
quasar/cluster selections. Rodrigues-William \& Hogan (1994)
and Seitz \& Schneider (1995) have already pointed out that the
patchy dust obscuration cannot explain their observations. 
Alternatively, the background quasar clustering is detected
only at $r<60$ $h_{50}^{-1}$ Mpc (Mo \& Fang 1993). This clustering
scale is comparable with the angular separation in the 
quasar-cluster associations but is 
much smaller than the spatial separation of the selected quasars.
Cluster-cluster autocorrelation seems to be  another possibility.
However, if the background quasars are detected randomly on the sky,
there would be no angular correlation between quasars and 
clusters even if the clusters are auto-correlated, which has been 
shown by Rodrigues-Williams \& Hogan (1994) using their data and also
by our simulations.  

Another way out of the difficulty is that there exists a large amount 
of unseen matter between clusters of galaxies on scale of 
less than $\sim20$ Mpc. Recall that we did not
include the contribution of the intercluster matter in the above
discussion. It is hard to figure out the distribution of this dark matter,
but it should be massive enough
to provide a surface density of as high as that
required by the quasar-cluster associations.  We will employ the 
N-body simulation of formation of clusters and large-scale
structures to further study the issue (Wu, Fang \& Jing, 1996)

It may be possible that the 
observed background quasar counts deviate significantly from
their intrinsic ones.  The fact that quasars are strongly associated
with the foreground galaxies and clusters indicates that the observations 
may preferentially  select those quasars whose angular positions appear to
be close to the foreground matter clumps. Unfortunately,
previous studies (Schneider
1987;1992; Pei 1995; references therein) about the magnification bias
on the observed quasar number counts reached a controversial result,
depending mainly on our current knowledge of
the distribution of lensing objects in the universe. 
It deserves to be investigated 
whether or not the quasar counts have been seriously contaminated by 
the lensing effect due to large-scale matter inhomogeneities. 
Meanwhile, our theoretical prediction of $q$ depends sensitively on
the adopted quasar counts (Boyle et al. 1988), which may have large
uncertainties. Recall that a different quasar number-magnitude
relation is derived by Hawkins \& V\'eron (1993).

Finally, 
we have tested the possibility of attributing the quasar-cluster
associations to the cluster environmental effect 
from the gravitational matter of cluster-galaxy correlation and 
will present the result elsewhere (Wu et al. 1996b).

\acknowledgments

We thank an anonymous referee for her/his valuable suggestions.
WXP was supported by the National Science Foundation
of China and a  World-Laboratory fellowship.

\clearpage

\begin{deluxetable}{cccccccccc}
\footnotesize
\tablecaption{Quasar-cluster associations: 
              observations and models. \label{table-1}}
\tablewidth{0pt}
\tablehead{
\colhead{clusters} & \colhead{quasars}   & \colhead{$\langle z_d\rangle^{a}$} 
 & \colhead{$\langle z_s\rangle^{b}$} & 
\colhead{$\theta^{c}$}  & 
\colhead{$\langle q\rangle_{obs}$} & 
\colhead{$(\sigma_v/10^3)^{d}$} &
\colhead{$\Sigma^{e}$} &
\colhead{$\Sigma^{f}$} &
\colhead{ref}
} 
\startdata
Zwicky & $B\leq18.5$ & 0.2 & 1.8 & 52 & $1.7_{-0.4}^{+0.5}$ & 
         $5.3_{-1.6}^{+1.6}$ &  
	 $0.10_{-0.05}^{+0.04}$ & $0.08_{-0.04}^{+0.04}$ & 1 \nl
Abell  & $S\geq2$ Jy & 0.1 & 2.0 & 24 & $1.7_{-0.5}^{+0.5}$ & 
         $4.7_{-1.8}^{+1.2}$ &  
	 $0.28_{-0.18}^{+0.10}$ & $0.25_{-0.16}^{+0.09}$ & 2 \nl
UKJ287$^{g}$ & $B\leq18.5$ & 0.15 & 1.5 & 7.2 & $2.0_{-0.2}^{+0.2}$ & 
         $2.3_{-0.2}^{+0.2}$ &  
	 $0.12_{-0.02}^{+0.02}$ & $0.11_{-0.02}^{+0.01}$ & 3 \nl
Zwicky & $\leq19$ & 0.2 & 1 & $78$ & $\sim1.3$ & 
         $4.3$ & $0.06$ & $0.05$ & 4 \nl
       & $\geq1$ Jy &  &  &  &  & 
         $5.6$ &  $0.11$ & $0.10$ &  \nl
\enddata

\tablenotetext{}{REFERENCES. -- (1)Rodrigues-Williams \& Hogan 1994;
		(2) Wu \& Han 1995; (3) Rodrigues \& Hawkins 1995;
                (4) Seitz \& Schneider 1995.}
\tablenotetext{a}{Mean cluster redshift}
\tablenotetext{b}{Mean quasar redshift}
\tablenotetext{c}{Search range in arcminutes}
\tablenotetext{d}{Required cluster velocity dispersion in 
	          units of 1000 km s$^{-1}$}
\tablenotetext{e}{Required surface mass density in g cm$^{-2}$ $h_{50}$
                 for $\Omega_0=1$ and $\lambda_0=0$}
\tablenotetext{f}{Required surface mass density in g cm$^{-2}$ $h_{50}$
                 for $\Omega_0=0.2$ and $\lambda_0=0.8$}
\tablenotetext{g}{Clusters in UKJ287 field}

\end{deluxetable}


\clearpage


\begin{references}
\reference{}Bartelmann, M. \& Schneider, P. 1993a, \aap, 268, 1
\reference{}Bartelmann, M. \& Schneider, P. 1993b, \aap, 271, 421
\reference{}Bartelmann, M. \& Schneider, P. 1994, \aap, 284, 1
\reference{}Boyle, R. J., Shanks, T., \& Peterson, B. A. 1988,
	  \mnras, 235, 935 
\reference{}Fang, L. Z. \& Wu, X. P. 1993, \apj, 408, 25
\reference{}Fugmann, W. 1988, \aap, 204, 73
\reference{}Fugmann, W. 1990, \aap, 240, 11
\reference{}Hawkins, M. R. S. \& Vv\'eron, P. 1993, \mnras, 260, 202
\reference{}Langston, G. I., Conner, S. R., Heflin, M. B., Leh\'ar, J.,
	    \& Burke, B. F. 1990, \apj, 353, 34
\reference{}Mo, H. J. \& Fang, L. Z. 1993, \apj, 410, 493
\reference{}Narayan, R. 1989, \apj, 339, L53
\reference{}Pei, Y. C. 1995, \apj, 440, 485
\reference{}Postman, M., Huchra, J. P., \& Geller, M. J. 1992, \apj, 
	    384, 404
\reference{}Rodrigues-Williams, L. L. \& Hogan, C. C. 1994,
	    \aj, 107, 451
\reference{}Rodrigues-Williams, L. L. \& Hawkins, M. R. S. 1995,
	    Proc. of the 5th Ann. Astrophys. Conf. (Maryland).
\reference{}Schneider, P. 1987, \apj,  316, 7
\reference{}Schneider, P. 1992, \aap,  254, 14
\reference{}Seitz, S. \& Schneider, P. 1995, \aap, 302, 9
\reference{}Turner, E. L., Ostriker, J. P., \& Gott III, J. R.
	  1984, \apj, 284, 1 
\reference{}Wu, X. P. 1994, \aap, 286, 748
\reference{}Wu, X. P., Fang, L. Z. \& Jing, Y. P. 1996, in preparation
\reference{}Wu, X. P., Fang, L. Z., Zhu, Z. \& Qin, B. 
	  1996, \apj, to be submitted
\reference{}Wu, X. P., Zhu, Z. H. \& Fang, L. Z. 1995, \apj, submitted 
\reference{}Wu, X. P. \& Han, J. 1995, \mnras, 272, 705
\end{references}
\end{document}